\documentclass[showpacs,showkeys,preprintnumbers]{revtex4}%
\usepackage{amsfonts}
\usepackage{amsmath}
\usepackage{graphicx}
\usepackage{dcolumn}
\usepackage{bm}
\usepackage{amssymb}%
\begin{document}
\title{Decay of the pseudoscalar glueball into scalar and pseudoscalar mesons}
\author{Walaa I.\ Eshraim$^{\text{(a)}}$, Stanislaus Janowski$^{\text{(a)}}$,
Francesco Giacosa$^{\text{(a)}}$, and Dirk H.\ Rischke$^{\text{(a,b)}}$}
\affiliation{$^{\text{(a)}}$Institute for Theoretical Physics, Goethe University,
Max-von-Laue-Str.\ 1, D--60438 Frankfurt am Main, Germany }
\affiliation{$^{\text{(b)}}$Frankfurt Institute for Advanced Studies, Goethe University,
Ruth-Moufang-Str.\ 1, D--60438 Frankfurt am Main, Germany }

\begin{abstract}
We study a chiral Lagrangian which describes the two- and three-body decays of
a pseudoscalar glueball into scalar and pseudoscalar mesons. The various
branching ratios are a parameter-free prediction of our approach. We compute
the decay channels for a pseudoscalar glueball with a mass of $2.6$ GeV, as
predicted by Lattice QCD in the quenched approximation, which is in the reach
of the PANDA experiment at the upcoming FAIR facility. For completeness, we
also repeat the calculation for a glueball mass of $2.37$ GeV which
corresponds to the mass of the resonance $X(2370)$ measured in the BESIII experiment.

\end{abstract}

\maketitle

\section{Introduction}

The fundamental symmetry underlying Quantum Chromodynamics (QCD), the theory
of strong interactions, is the exact local $SU(3)_{c}$ color symmetry. As a
consequence of the non-abelian nature of this symmetry the gauge fields of
QCD, the gluons, are colored objects and therefore interact strongly with each
other. Because of confinement, one expects that gluons can also form
colorless, or `white', states which are called glueballs.

The first calculations of glueball masses were based on the bag-model approach
\cite{bag-glueball}. Later on, the rapid improvement of lattice QCD allowed
for precise simulations of Yang-Mills theory, leading to a determination of
the full glueball spectrum \cite{Morningstar}. However, in full QCD (i.e.,
gluons plus quarks) the mixing of glueball and quark-antiquark configurations
with the same quantum number occurs, rendering the identification of the
resonances listed in the Particle Data Group (PDG) \cite{PDG} more difficult.
The search for states which are (predominantly) glueball represents an active
experimental and theoretical area of research, see Ref.\ \cite{review} and
refs.\ therein. The reason for these efforts is that a better understanding of
the glueball properties would represent an important step in the comprehension
of the non-perturbative behavior of QCD. However, although up to now some
glueball candidates exist (see below), no state which is (predominantly)
glueball has been unambiguously identified.

In general, a glueball state should fulfill two properties regarding its
decays: it exhibits `flavor blindness', because the gluons couple with the
same strength to all quark flavors, and it is narrow, because QCD in the
large-$N_{c}$ limit shows that all glueball decay widths scale as $N_{c}%
^{-2},$ which should be compared to the $N_{c}^{-1}$ scaling law for a
quark-antiquark state. The lightest glueball state predicted by lattice QCD
simulations is a scalar-isoscalar state ($J^{PC}=0^{++}$) with a mass of about
$1.7$ GeV \cite{Morningstar}. The resonance $f_{0}(1500)$ shows a flavor-blind
decay pattern and is narrow, thus representing a good candidate for a state
which is (predominantly) a scalar glueball. Also the resonance $f_{0}(1710)$
is a glueball candidate because its mass is very close to lattice QCD
predictions and it is copiously produced in the gluon-rich decay of the
$J/\psi$ meson. Both scenarios have been investigated in a variety of works,
e.g.\ Refs.\ \cite{scalars,giacosa,cheng,stani,gutsche} and refs.\ therein, in
which mixing patterns involving the scalar resonances $f_{0}(1370),$
$f_{0}(1500)$, and $f_{0}(1710)$ are considered. In particular, in
Ref.\ \cite{gutsche} the decays of the $J/\psi$ have been included in a
phenomenological fit and both assignments turn out to be consistent, but
slightly favour a predominant gluonic amount in $f_{0}(1500)$.

The second lightest lattice-predicted glueball state has tensor quantum
numbers ($J^{PC}=2^{++}$) and a mass of about $2.2$ GeV; a good candidate
could be the very narrow resonance $f_{J}(2200)$ \cite{tensor,burakovsky}, if
the total spin of the latter will be experimentally confirmed to be $J=2$.

The third least massive glueball predicted by lattice QCD (in the quenched
approximation) has pseudoscalar quantum numbers ($J^{PC}=0^{-+}$) and a mass
of about $2.6$ GeV. Quite remarkably, most theoretical works investigating the
pseudoscalar glueball did not take into account this prediction of Yang-Mills
lattice studies, but concentrated their search around $1.5$ GeV in connection
with the isoscalar-pseudoscalar resonances $\eta(1295),\eta(1405)$, and
$\eta(1475)$. A candidate for a predominantly light pseudoscalar glueball is
the middle-lying state $\eta(1405)$ due to the fact that it is largely
produced in (gluon-rich) $J/\psi$ radiative decays and is missing in
$\gamma\gamma$ reactions \cite{mixetas}. In this framework the resonances
$\eta(1295)$ and $\eta(1475)$ represent radial excitations of the resonances
$\eta$ and $\eta^{\prime}$. Indeed, in relation to $\eta$ and $\eta^{\prime}$,
a lot of work has been done in determining the gluonic amount of their wave
functions. The KLOE Collaboration found that the pseudoscalar glueball
fraction in the mixing of the pseudoscalar-isoscalar states $\eta$ and
$\eta^{\prime}$ can be large ($\sim14$\%) \cite{KLOEpsglue}, but the
theoretical work of Ref.\ \cite{escribano} found that the glueball amount in
$\eta$ and $\eta^{\prime}$ is compatible with zero [see, however, also
Ref.\ \cite{escribanonew}]\textbf{.}

In this work we study the decay properties of a pseudoscalar glueball state
whose mass lies, in agreement with lattice QCD, between $2$ and $3$ GeV.
Following Ref.\ \cite{schechter} we write down an effective chiral Lagrangian
which couples the pseudoscalar glueball field (denoted as $\tilde{G}$) to
scalar and pseudoscalar mesons. We can thus evaluate the widths for the decays
$\tilde{G}\rightarrow PPP$ and $\tilde{G}\rightarrow PS,$ where $P$ and $S$
stand for pseudoscalar and scalar quark-antiquark states. The pseudoscalar
state $P$ refers to the well-known light pseudoscalars $\{\pi,K,\eta
,\eta^{\prime}\}$, while the scalar state $S$ refers to the quark-antiquark
nonet of scalars above 1 GeV: $\{a_{0}(1450),K_{0}^{\ast}(1430),f_{0}%
(1370),f_{0}(1500)\,\mathrm{or}\,f_{0}(1710)\}$. The reason for the latter
assignment is a growing consensus that the chiral partners of the pseudoscalar
states should not be identified with the resonances below 1 GeV, see
Refs.\ \cite{denis,stani,nf3} for results within the so-called extended linear
sigma model and also other theoretical works in Ref.\ \cite{scalars,varietq}
(and refs.\ therein).

The chiral Lagrangian that we construct contains one unknown coupling constant
which cannot be determined without experimental data. However, the branching
ratios can be unambiguously calculated and may represent a useful guideline
for experimental search of the pseudoscalar glueball in the energy region
between $2$ to $3$ GeV. In this respect, the planned PANDA experiment at the
FAIR facility \cite{panda} will be capable to scan the mass region above 2.5
GeV. The experiment is based on proton-antiproton scattering, thus the
pseudoscalar glueball $\tilde{G}$ can be directly produced as an intermediate
state. We shall therefore present our results for the branching ratios for a
putative pseudoscalar glueball with a mass of 2.6 GeV.

On the other hand, it is also possible that the pseudoscalar glueball
$\tilde{G}$ has a mass that is a bit lower than the lattice QCD prediction and
that it has been already observed in the BESIII experiment where pseudoscalar
resonances have been investigated in $J/\psi$ decays \cite{bes}. In
particular, the resonance $X(2370)$ which has been clearly observed in the
$\pi^{+}\pi^{-}\eta^{\prime}$ channel represents a good candidate, because it
is quite narrow ($\sim80$ MeV) and its mass lies just below the lattice QCD
prediction. For this reason we repeat our calculation for a pseudoscalar
glueball mass of $2.37$ GeV, and thus make predictions for the resonance
$X(2370)$, which can be tested in the near future.

This paper is organized as follows. In Sec.\ II we present the effective
Lagrangian coupling the pseudoscalar glueball to scalar and pseudoscalar
quark-antiquark degrees of freedom, and we calculate the branching ratios for
the decays into $PPP$ and $SP$. Finally, in Sec.\ III we present our
conclusions and an outlook.

\section{The effective Lagrangian}

Following Ref.\ \cite{schechter} we introduce a chiral Lagrangian which
couples the pseudoscalar glueball $\tilde{G}\equiv\left\vert gg\right\rangle $
with quantum numbers $J^{PC}=0^{-+}$ to scalar and pseudoscalar mesons
\begin{equation}
\mathcal{L}_{\tilde{G}}^{int}=ic_{\tilde{G}\Phi}\tilde{G}\left(
\text{\textrm{det}}\Phi-\text{\textrm{det}}\Phi^{\dag}\right)  \text{ ,}
\label{intlag}%
\end{equation}
where $c_{\tilde{G}\Phi}$ is a coupling constant,
\begin{equation}
\Phi=(S^{a}+iP^{a})t^{a} \label{phimat}%
\end{equation}
represents the multiplet of scalar and pseudoscalar quark-antiquark states,
and $t^{a}$ are the generators of the group $U(N_{f})$. In this work we
consider the case $N_{f}=3$ and the explicit representation of the scalar and
pseudoscalar mesons reads \cite{nf3,dick}:%
\begin{equation}
\Phi=\frac{1}{\sqrt{2}}\left(
\begin{array}
[c]{ccc}%
\frac{(\sigma_{N}+a_{0}^{0})+i(\eta_{N}+\pi^{0})}{\sqrt{2}} & a_{0}^{+}%
+i\pi^{+} & K_{S}^{+}+iK^{+}\\
a_{0}^{-}+i\pi^{-} & \frac{(\sigma_{N}-a_{0}^{0})+i(\eta_{N}-\pi^{0})}%
{\sqrt{2}} & K_{S}^{0}+iK^{0}\\
K_{S}^{-}+iK^{-} & \bar{K}_{S}^{0}+i\bar{K}^{0} & \sigma_{S}+i\eta_{S}%
\end{array}
\right)  \;. \label{phimatex}%
\end{equation}
Under $U_{L}(3)\times U_{R}(3)$ chiral transformations the multiplet $\Phi$
transforms as $\Phi\rightarrow U_{L}\Phi U_{R}^{\dagger}$ where $U_{L}$ and
$U_{R}$ are $U(3)$ matrices. The determinant of $\Phi$ is invariant under
$SU(3)_{L}\times SU(3)_{R}$, but not under $U(1)_{A}$. On the other hand, the
pseudoscalar glueball field $\tilde{G}$ is invariant under $U(3)_{L}\times
U(3)_{R}$ transformations. Under parity, $\Phi\rightarrow\Phi^{\dagger}$ and
$\tilde{G}\rightarrow-\tilde{G}$, thus the effective Lagrangian of
Eq.\ (\ref{intlag}) is invariant under $SU(3)_{L}\times SU(3)_{R}$ and under
parity. Notice that Eq.\ (\ref{intlag}) is not invariant under $U_{A}(1)$, in
agreement with the so-called axial anomaly in the isoscalar-pseudoscalar
sector. The rest of the mesonic Lagrangian which describes the interactions of
$\Phi$ and also includes (axial-)vector degrees of freedom is presented in
Sec.\ \ref{app1} of the Appendix. For more details, see
Refs.\ \cite{dick,nf3,denisthesis}.

The assignment of the quark-antiquark fields in this paper is as follows: (i)
In the pseudoscalar sector the fields $\vec{\pi}$ and $K$ represent the pions
or the kaons, respectively \cite{PDG}. The bare fields $\eta_{N}%
\equiv\left\vert \bar{u}u+\bar{d}d\right\rangle /\sqrt{2}$ and $\eta_{S}%
\equiv\left\vert \bar{s}s\right\rangle $ are the non-strange and strange
contributions of the physical states $\eta$ and $\eta^{\prime}$ \cite{PDG}:%
\begin{equation}
\eta=\eta_{N}\cos\varphi+\eta_{S}\sin\varphi,\text{ }\eta^{\prime}=-\eta
_{N}\sin\varphi+\eta_{S}\cos\varphi, \label{mixetas}%
\end{equation}
where $\varphi\simeq-44.6^{\circ}$ is the mixing angle \cite{dick}. Using
other values for the mixing angle, e.g.\ $\varphi=-36^{\circ}$
\cite{Francescomixetas} or $\varphi=-41.4^{\circ}$, as determined by the KLOE
Collaboration \cite{KLOEpsglue}, affects the presented results only
marginally. (ii) In the scalar sector we assign the field $\vec{a}_{0}$ to the
physical isotriplet state $a_{0}(1450)$ and the scalar kaon fields $K_{S}$ to
the resonance $K_{0}^{\star}(1430).$ As a first approximation, the non-strange
bare field $\sigma_{N}\equiv\left\vert \bar{u}u+\bar{d}d\right\rangle
/\sqrt{2}$ is assigned to the physical isoscalar resonance \thinspace
$f_{0}(1370)$ and the bare field $\sigma_{S}\equiv\left\vert \bar
{s}s\right\rangle $ is assigned either to $f_{0}(1710)$ or to $f_{0}(1500).$
In a more complete framework, $\sigma_{N},$ $\sigma_{S}$ and a bare scalar
glueball field $G$ mix and generate the physical resonances $f_{0}(1370),$
$f_{0}(1500)$, and $f_{0}(1710),$ see the discussion below.

In order to evaluate the decays of the pseudoscalar glueball $\tilde{G}$ we
have to take into account that the spontaneous breaking of chiral symmetry
takes place, which implies the need of shifting the scalar-isoscalar fields by
their vacuum expectation values $\phi_{N}$ and $\phi_{S}$,
\begin{equation}
\sigma_{N}\rightarrow\sigma_{N}+\phi_{N}\text{ and }\sigma_{S}\rightarrow
\sigma_{S}+\phi_{S}\text{ .} \label{shift}%
\end{equation}
In addition, when (axial-)vector mesons are present in the Lagrangian, one
also has to `shift' the axial-vector fields and to define the wave-function
renormalization constants of the pseudoscalar fields:%
\begin{equation}
\vec{\pi}\rightarrow Z_{\pi}\vec{\pi}\text{ , }K^{i}\rightarrow Z_{K}%
K^{i}\text{, }\eta_{j}\rightarrow Z_{\eta_{j}}\eta_{j}\;, \label{psz}%
\end{equation}
where $i=1,2,3,4$ runs over the four kaonic fields and $j=N,S.$ The numerical
values of the renormalization constants are $Z_{\pi}=1.709$, $Z_{K}%
=1.604,Z_{K_{S}}=1.001,$ $Z_{\eta_{N}}=Z_{\pi},$ $Z_{\eta_{S}}=1.539$
\cite{dick}. Moreover, the condensates $\phi_{N}$ and $\phi_{S}$ read%
\begin{equation}
\phi_{N}=Z_{\pi}f_{\pi}=0.158\text{ GeV, }\phi_{S}=\frac{2Z_{K}f_{K}-\phi_{N}%
}{\sqrt{2}}=0.138\text{ GeV}\;,
\end{equation}
where the standard values $f_{\pi}=0.0922$ GeV and $f_{K}=0.110$ GeV have been
used \cite{PDG}. Once the operations in Eqs.\ (\ref{shift}) and (\ref{psz})
have been performed, the Lagrangian in Eq.\ (\ref{intlag}) contains the
relevant tree-level vertices for the decay processes of $\tilde{G}$, see
Appendix (Sec.\ \ref{app2}).

The branching ratios of $\tilde{G}$ for the decays into three pseudoscalar
mesons are reported in Table I for both choices of the pseudoscalar masses,
$2.6$ and $2.37$ GeV (relevant for PANDA and BESIII experiments,
respectively). The branching ratios are presented relative to the total decay
width of the pseudoscalar glueball $\Gamma_{\tilde{G}}^{tot}$. (For details of
the calculation of the three-body decay we refer to Sec.\ \ref{app3} of the Appendix.)

\begin{center}%
\begin{table}[h] \centering
\begin{tabular}
[c]{|c|c|c|}\hline
Quantity & Case (i): $M_{\tilde{G}}=2.6$ GeV & Case (ii): $M_{\tilde{G}}=2.37$
GeV\\\hline
$\Gamma_{\tilde{G}\rightarrow KK\eta}/\Gamma_{\tilde{G}}^{tot}$ & $0.049$ &
$0.043$\\\hline
$\Gamma_{\tilde{G}\rightarrow KK\eta^{\prime}}/\Gamma_{\tilde{G}}^{tot}$ &
$0.019$ & $0.011$\\\hline
$\Gamma_{\tilde{G}\rightarrow\eta\eta\eta}/\Gamma_{\tilde{G}}^{tot}$ & $0.016$
& $0.013$\\\hline
$\Gamma_{\tilde{G}\rightarrow\eta\eta\eta^{\prime}}/\Gamma_{\tilde{G}}^{tot}$
& $0.0017$ & $0.00082$\\\hline
$\Gamma_{\tilde{G}\rightarrow\eta\eta^{\prime}\eta^{\prime}}/\Gamma_{\tilde
{G}}^{tot}$ & $0.00013$ & $0$\\\hline
$\Gamma_{\tilde{G}\rightarrow KK\pi}/\Gamma_{\tilde{G}}^{tot}$ & $0.47$ &
$0.47$\\\hline
$\Gamma_{\tilde{G}\rightarrow\eta\pi\pi}/\Gamma_{\tilde{G}}^{tot}$ & $0.16$ &
$0.17$\\\hline
$\Gamma_{\tilde{G}\rightarrow\eta^{\prime}\pi\pi}/\Gamma_{\tilde{G}}^{tot}$ &
$0.095$ & $0.090$\\\hline
\end{tabular}%
\caption{Branching ratios for the decay of the pseudoscalar glueball $\tilde
{G}$ into three pseudoscalar mesons.}%
\end{table}%

\end{center}

Next we turn to the decay process $\tilde{G}\rightarrow PS.$ The results, for
both choices of $M_{\tilde{G}},$ are reported in Table II for the cases in
which the bare resonance $\sigma_{S}$ is assigned to $f_{0}(1710)$ or to
$f_{0}(1500).$

\begin{center}%
\begin{table}[h] \centering
\begin{tabular}
[c]{|c|c|c|}\hline
Quantity & Case (i): $M_{\tilde{G}}=2.6$ GeV & Case (ii): $M_{\tilde{G}}=2.37$
GeV\\\hline
$\Gamma_{\tilde{G}\rightarrow KK_{S}}/\Gamma_{\tilde{G}}^{tot}$ & $0.060$ &
$0.070$\\\hline
$\Gamma_{\tilde{G}\rightarrow a_{0}\pi}/\Gamma_{\tilde{G}}^{tot}$ & $0.083$ &
$0.10$\\\hline
$\Gamma_{\tilde{G}\rightarrow\eta\sigma_{N}}/\Gamma_{\tilde{G}}^{tot}$ &
$0.0000026$ & $0.0000030$\\\hline
$\Gamma_{\tilde{G}\rightarrow\eta^{\prime}\sigma_{N}}/\Gamma_{\tilde{G}}%
^{tot}$ & $0.039$ & $0.026$\\\hline
$\Gamma_{\tilde{G}\rightarrow\eta\sigma_{S}}/\Gamma_{\tilde{G}}^{tot}$ &
$0.012$ $(0.015)$ & $0.0094$ $(0.017)$\\\hline
$\Gamma_{\tilde{G}\rightarrow\eta^{\prime}\sigma_{S}}/\Gamma_{\tilde{G}}%
^{tot}$ & $0$ $(0.0082)$ & $0$ $(0)$\\\hline
\end{tabular}%
\caption{Branching ratios for the decay of the pseudoscalar glueball $\tilde
{G}$ into a scalar and a pseudoscalar meson. In the last two rows
$\sigma_{S}$ is assigned to $f_{0}(1710)$ or to $f_{0}(1500)$
(values in the parentheses).}%
\end{table}%

\end{center}

Concerning the decays involving scalar-isoscalar mesons, one should go beyond
the results of Table II by including the full mixing pattern above 1 GeV, in
which the resonances $f_{0}(1370),$ $f_{0}(1500),$ and $f_{0}(1710)$ are mixed
states of the bare quark-antiquark contributions $\sigma_{N}\equiv\left\vert
\bar{u}u+\bar{d}d\right\rangle /\sqrt{2}$ and $\sigma_{S}$ and a bare scalar
glueball field $G.$ This mixing is described by an orthogonal $(3\times3)$
matrix \cite{scalars,giacosa,cheng,stani,gutsche}. In view of the fact that a
complete evaluation of this mixing in the framework of our chiral approach has
not yet been done, we use the two solutions for the mixing matrix of
Ref.\ \cite{giacosa} and the solution of Ref.\ \cite{cheng} in order to
evaluate the decays of the pseudoscalar glueball into the three
scalar-isoscalar resonances $f_{0}(1370),$ $f_{0}(1500),$ and $f_{0}(1710)$.
In all three solutions $f_{0}(1370)$ is predominantly described by the bare
configuration $\sigma_{N}\equiv\left\vert \bar{u}u+\bar{d}d\right\rangle
/\sqrt{2}$, but the assignments for the other resonances vary: in the first
solution of Ref.\ \cite{giacosa} the resonance $f_{0}(1500)$ is predominantly
gluonic, while in the second solution of Ref.\ \cite{giacosa} and the solution
of Ref.\ \cite{cheng} the resonance $f_{0}(1710)$ has the largest gluonic
content. The results for the decay of the pseudoscalar glueball into
scalar-isoscalar resonances are reported in Table III.%

\begin{table}[h] \centering
\begin{tabular}
[c]{|c|c|c|c|}\hline
Quantity & Sol. 1 of Ref. \cite{giacosa} & Sol. 2 of Ref. \cite{giacosa} &
Sol. of Ref. \cite{cheng}\\\hline
$\Gamma_{\tilde{G}\rightarrow\eta f_{0}(1370)}/\Gamma_{\tilde{G}}^{tot}$ &
\multicolumn{1}{|l|}{$0.00093$ $(0.0011)$} & \multicolumn{1}{|l|}{$0.00058$
$(0.00068)$} & \multicolumn{1}{|l|}{$0.0044$ $(0.0052)$}\\\hline
$\Gamma_{\tilde{G}\rightarrow\eta f_{0}(1500)}/\Gamma_{\tilde{G}}^{tot}$ &
\multicolumn{1}{|l|}{$0.000046$ $(0.000051)$} & \multicolumn{1}{|l|}{$0.0082$
$(0.0090)$} & \multicolumn{1}{|l|}{$0.011$ $(0.012)$}\\\hline
$\Gamma_{\tilde{G}\rightarrow\eta f_{0}(1710)}/\Gamma_{\tilde{G}}^{tot}$ &
\multicolumn{1}{|l|}{$0.011$ $(0.0089)$} & \multicolumn{1}{|l|}{$0.0053$
$(0.0042)$} & \multicolumn{1}{|l|}{$0.00037$ $(0.00029)$}\\\hline
$\Gamma_{\tilde{G}\rightarrow\eta^{\prime}f_{0}(1370)}/\Gamma_{\tilde{G}%
}^{tot}$ & \multicolumn{1}{|l|}{$0.038$ $(0.026)$} &
\multicolumn{1}{|l|}{$0.033$ $(0.022)$} & \multicolumn{1}{|l|}{$0.043$
$(0.029)$}\\\hline
$\Gamma_{\tilde{G}\rightarrow\eta^{\prime}f_{0}(1500)}/\Gamma_{\tilde{G}%
}^{tot}$ & \multicolumn{1}{|l|}{$0.0062$ $(0)$} &
\multicolumn{1}{|l|}{$0.00020$ $(0)$} & \multicolumn{1}{|l|}{$0.00013$ $(0)$%
}\\\hline
$\Gamma_{\tilde{G}\rightarrow\eta^{\prime}f_{0}(1710)}/\Gamma_{\tilde{G}%
}^{tot}$ & \multicolumn{1}{|l|}{$0$ $(0)$} & \multicolumn{1}{|l|}{$0$ $(0)$} &
\multicolumn{1}{|l|}{$0$ $(0)$}\\\hline
\end{tabular}%
\caption{Branching ratios for the decays of the pseudoscalar glueball $\tilde
{G}$ into $\eta$ and $\eta
'$, respectively and one of the scalar-isoscalar states: $f_0(1370), f_0(1500)$ and $ f_0(1710)$
by using three different mixing scenarios of these scalar-isoscalar states reported in Refs \cite
{giacosa,cheng}.
The mass of the pseudoscalar glueball is $M_{\tilde{G}}%
=2.6$ GeV and $M_{\tilde{G}}=2.37$ GeV (values  in the brackets),
respectively.}%
\end{table}%

\begin{figure}
[ptb]
\begin{center}
\includegraphics[
height=2.1958in,
width=3.1099in
]%
{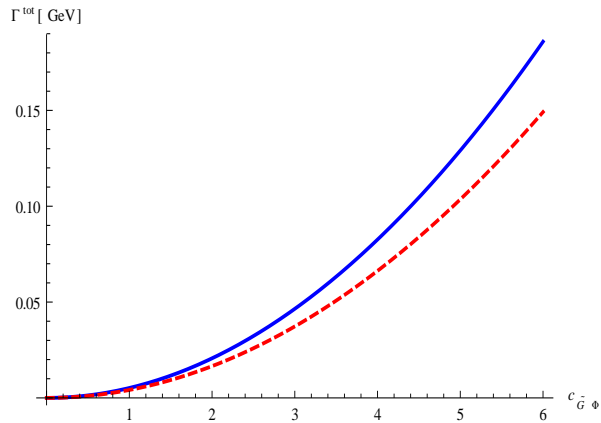}%
\caption{{\small Solid (blue) line: Total decay width of the pseudoscalar
glueball with the bare mass }${\small M}_{\tilde{G}}{\small =2.6}${\small
\ GeV as function of the coupling }$c_{\tilde{G}\Phi}${\small  . Dashed (red)
line: Same curve for }${\small M}_{\tilde{G}}{\small =2.37}$ {\small GeV.}}%
\label{fig1}%
\end{center}
\end{figure}

In Fig.\ \ref{fig1} we show the behavior of the total decay width
$\Gamma_{\tilde{G}}^{tot}=\Gamma_{\tilde{G}\rightarrow PPP}+\Gamma_{\tilde
{G}\rightarrow PS}$ as function of the coupling constant $c_{\tilde{G}\Phi}$
for both choices of the pseudoscalar glueball mass. (We assume here that other
decay channels, such as decays into vector mesons or baryons are negligible).
In the case of $M_{\tilde{G}}=2.6$ GeV, one expects from large-$N_{c}$
considerations that the total decay width $\Gamma_{\tilde{G}}^{tot}%
\lesssim100$ MeV. In fact, as discussed in the Introduction, the scalar
glueball candidate $f_{0}(1500)$ is roughly $100$ MeV broad and the tensor
candidate $f_{J}(2220)$ is even narrower. In the present work, the condition
$\Gamma_{\tilde{G}}^{tot}\lesssim100$ MeV implies that $c_{\tilde{G}\Phi
}\lesssim$ $5$. Moreover, in the case of $M_{\tilde{G}}=2.37$ GeV in which the
identification $\tilde{G}\equiv X(2370)$ has been made, we can indeed use the
experimental knowledge on the full decay width [$\Gamma_{X(2370)}=83\pm17$ MeV
\cite{bes}] to determine the coupling constant to be $c_{\tilde{G}\Phi
}=4.48\pm0.46$. (However, we also refer to the recent work of
Ref.\ \cite{bugg}, where the possibility of a broad pseudoscalar glueball is discussed.)%

Some comments are in order:

(i) The results depend only slightly on the glueball mass, thus the two
columns of Table I and II are similar. It turns out that the channel $KK\pi$
is the dominant one (almost 50\%). Also the $\eta\pi\pi$ and $\eta^{\prime}%
\pi\pi$ channels are sizable. On the contrary, the two-body decays are
subdominant and reach only 20\% of the full mesonic decay width.

(ii) The decay of the pseudoscalar glueball into three pions vanishes:
\begin{equation}
\Gamma_{\tilde{G}\rightarrow\pi\pi\pi}=0\text{ .}%
\end{equation}
This result represents a further testable prediction of our approach.

(iii) The decays of the pseudoscalar glueball into a scalar-isoscalar meson
amount only to $5\%$ of the total decay width. Moreover, the mixing pattern in
the scalar-isoscalar sector has a negligible influence on the total decay
width of $\tilde{G}.$ Nevertheless, in the future it may represent an
interesting and additional test for scalar-isoscalar states.

(iv) Once the shifts of the scalar fields have been performed, there are also
bilinear mixing terms of the form $\tilde{G}\eta_{N}$ and $\tilde{G}\eta_{S}$
which lead to a non-diagonal mass matrix. In principle, one should take these
terms into account, in addition to the already mentioned $\eta_{N}\eta_{S}$
mixing, and solve a three-state mixing problem in order to determine the
masses of the pseudoscalar particles. This will also affect the calculation of
the decay widths. However, due to the large mass difference of the bare
glueball fields $\tilde{G}$ to the other quark-antiquark pseudoscalar fields,
the mixing of $\tilde{G}$ turns out to be very small in the present work, and
can be safely neglected. For instance, it turns out that the mass of the mixed
state which is predominantly glueball is (at most) just $0.002$ GeV larger
than the bare mass $M_{\tilde{G}}=2.6$ GeV.

(v) If a standard linear sigma model without (axial-)vector mesons is studied,
the replacements $Z_{\pi}=Z_{K}=Z_{\eta_{N}}=Z_{\eta_{S}}=1$ need to be
performed. Most of the results of the branching ratios for the three-body
decay are qualitatively, but not quantitatively, similar to the values of
Table I (variations of about $25$-$30\%$). However, the branching ratios for
the two-body decay change sizably w.r.t. the results of Table 2. This fact
shows once more that the inclusion of (axial-)vector degrees of freedom has
sizable effects also concerning the decays of the pseudoscalar glueball.

(vi) In principle, the three-body final states for the decays shown in Table I
can also be reached through a sequential decay from the two-body final states
shown in Table II, where the scalar particle $S$ further decays into $PP$, for
instance, $K_{0}^{\ast}(1430)\rightarrow K\pi$. There are then two possible
decay amplitudes, one from the direct three-body decay and one from the
sequential decay, which have to be added coherently before taking the modulus
square to obtain the total three-body decay width. Summing the results shown
in Table I and II gives a first estimate (which neglects interference terms)
for the magnitude of the total three-body decay width. We have verified that
the correction from the interference term to this total three-body decay width
in a given channel is at most of the order of $10$\% for $M_{\tilde{G}}=2.6$
GeV and $15\%$ for $M_{\tilde{G}}=2.37$ GeV. For a full understanding of the
contribution of the various decay amplitudes to the final three-body state,
one needs to perform a detailed study of the Dalitz plot for the three-body decay.

\section{Conclusions and Outlook}

In this work we have presented a chirally invariant effective Lagrangian
describing the interaction of the pseudoscalar glueball with scalar and
pseudoscalar mesons for the three-flavor case $N_{f}=3$. We have studied the
decays of the pseudoscalar glueball into three pseudoscalar and into a scalar
and pseudoscalar quark-antiquark fields.

The branching ratios are parameter-free once the mass of the glueball has been
fixed. We have considered two possibilities: (i) in agreement with lattice QCD
in the quenched approximation we have chosen $M_{\tilde{G}}=2.6$ GeV. The
existence and the decay properties of such a hypothetical pseudoscalar
resonance can be tested in the upcoming PANDA experiment \cite{panda}. (ii) We
assumed that the resonance $X(2370),$ measured in the experiment BESIII, is
(predominantly) a pseudoscalar glueball state, and thus we have also used a
mass of $2.37$ GeV \cite{bes}. The results for both possibilities have been
summarized in Tables I and II: we predict that $KK\pi$ is the dominant decay
channel, followed by (almost equally large) $\eta\pi\pi$ and $\eta^{\prime}%
\pi\pi$ decay channels. On the contrary, the decay into three pions is
predicted to vanish. In the case of BESIII, by measuring the branching ratio
for other decay channels than the measured $\eta^{\prime}\pi\pi$, one could
ascertain if $X(2370)$ is (predominantly) a pseudoscalar glueball. In the case
of PANDA, our results may represent a useful guideline for the search of the
pseudoscalar glueball.

Future studies should consider possible mixing of the pseudoscalar glueball
with charmonia states\textbf{ }and an improved description of the
scalar-isoscalar sector. New lattice results for the pseudoscalar glueball
mass, which go beyond the quenched approximation and include the effect of
dynamical fermions, would be very useful for model building. Moreover, the
mechanism of the glueball production via proton-antiproton fusion using the
so-called mirror assignment \cite{gallas} represents an interesting outlook
\cite{acta}.

\section*{Acknowledgments}

The authors thank Klaus Neuschwander for computing the interference terms for
the sequential decay into three-body final states. They also thank Diego
Bettoni, Stephen Olsen, Denis Parganlija, Anja Habersetzer, Antje Peters,
Klaus Neuschwander, and Marc Wagner for useful discussions. W.E.\ acknowledges
support from DAAD and HGS-HIRe, S.J.\ acknowledges support from H-QM and
HGS-HIRe. F.G.\ thanks the Foundation Polytechnical Society Frankfurt am Main
for support through an Educator fellowship.

\appendix

\section{Details of the calculation}

\subsection{The full mesonic Lagrangian}

\label{app1}

The chirally invariant $U(N_{f})_{L}\times U(N_{f})_{R}$ Lagrangian for the
low-lying mesonic states with (pseudo)scalar and (axial-)vector quantum
numbers has the form
\begin{align}
\mathcal{L}_{mes}  &  =\mathrm{Tr}[(D_{\mu}\Phi)^{\dagger}(D^{\mu}\Phi
)]-m_{0}^{2}\mathrm{Tr}(\Phi^{\dagger}\Phi)-\lambda_{1}[\mathrm{Tr}%
(\Phi^{\dagger}\Phi)]^{2}-\lambda_{2}\mathrm{Tr}(\Phi^{\dagger}\Phi
)^{2}\nonumber\\
&  -\frac{1}{4}\mathrm{Tr}[(L^{\mu\nu})^{2}+(R^{\mu\nu})^{2}]+\mathrm{Tr}%
[(\frac{m_{1}^{2}}{2}+\Delta)(L_{\mu}^{2}+R_{\mu}^{2})]+\mathrm{Tr}%
[H(\Phi+\Phi^{\dagger})]\nonumber\\
&  +c_{1}(\mathrm{det}\Phi-\mathrm{det}\Phi^{\dagger})^{2}+i\frac{g_{2}}%
{2}\{\mathrm{Tr}(L_{\mu\nu}[L^{\mu},L^{\nu}])+\mathrm{Tr}(R_{\mu\nu}[R^{\mu
},R^{\nu}])\}\nonumber\\
&  +\frac{h_{1}}{2}\mathrm{Tr}(\Phi^{\dagger}\Phi)\mathrm{Tr}\left(  L_{\mu
}^{2}+R_{\mu}^{2}\right)  +h_{2}\mathrm{Tr}[\left\vert L_{\mu}\Phi\right\vert
^{2}+\left\vert \Phi R_{\mu}\right\vert ^{2}]\nonumber\\
&  +2h_{3}\mathrm{Tr}(L_{\mu}\Phi R^{\mu}\Phi^{\dagger}). \label{fulllag}%
\end{align}
where
\[
L_{\mu}=\frac{1}{\sqrt{2}}\left(
\begin{array}
[c]{ccc}%
\frac{\omega_{N}^{\mu}+\rho^{\mu0}}{\sqrt{2}}+\frac{f_{1N}^{\mu}+a_{1}^{\mu0}%
}{\sqrt{2}} & \rho^{\mu+}+a_{1}^{\mu+} & K^{\star\mu+}+K_{1}^{\mu+}\\
\rho^{\mu-}+a_{1}^{\mu-} & \frac{\omega_{N}^{\mu}-\rho^{\mu0}}{\sqrt{2}}%
+\frac{f_{1N}^{\mu}-a_{1}^{\mu0}}{\sqrt{2}} & K^{\star\mu0}+K_{1}^{\mu0}\\
K^{\star\mu-}+K_{1}^{\mu-} & \bar{K}^{\star\mu0}+\bar{K}_{1}^{\mu0} &
\omega_{S}^{\mu}+f_{1S}^{\mu}%
\end{array}
\right)  \;,
\]
and
\[
R_{\mu}=\frac{1}{\sqrt{2}}\left(
\begin{array}
[c]{ccc}%
\frac{\omega_{N}^{\mu}+\rho^{\mu0}}{\sqrt{2}}-\frac{f_{1N}^{\mu}+a_{1}^{\mu0}%
}{\sqrt{2}} & \rho^{\mu+}-a_{1}^{\mu+} & K^{\star\mu+}-K_{1}^{\mu+}\\
\rho^{\mu-}-a_{1}^{\mu-} & \frac{\omega_{N}^{\mu}-\rho^{\mu0}}{\sqrt{2}}%
-\frac{f_{1N}^{\mu}-a_{1}^{\mu0}}{\sqrt{2}} & K^{\star\mu0}-K_{1}^{\mu0}\\
K^{\star\mu-}-K_{1}^{\mu-} & \bar{K}^{\star\mu0}-\bar{K}_{1}^{\mu0} &
\omega_{S}^{\mu}-f_{1S}^{\mu}%
\end{array}
\right)  \;,
\]
for details see Refs.\ \cite{dick,nf3,denisthesis,stani}. In the present
context we are interested in the wave-function renormalization constants
$Z_{i}$ introduced in Eq.\ (\ref{psz}). Their explicit expressions read
\cite{dick,denisthesis}:%
\begin{equation}
Z_{\pi}=Z_{\eta_{N}}=\frac{m_{a_{1}}}{\sqrt{m_{a_{1}}^{2}-g_{1}^{2}\phi
_{N}^{2}}}\text{ , }Z_{K}=\frac{2m_{K_{1}}}{\sqrt{4m_{K_{1}}^{2}-g_{1}%
^{2}(\phi_{N}+\sqrt{2}\phi_{S})^{2}}}\text{ ,} \label{zpi}%
\end{equation}%
\begin{equation}
Z_{K_{S}}=\frac{2m_{K^{\star}}}{\sqrt{4m_{K^{\star}}^{2}-g_{1}^{2}(\phi
_{N}-\sqrt{2}\phi_{S})^{2}}}\text{ , }Z_{\eta_{S}}=\frac{m_{f_{1S}}}%
{\sqrt{m_{f_{1S}}^{2}-2g_{1}^{2}\phi_{S}^{2}}}\;\text{.} \label{zets}%
\end{equation}

\subsection{Explicit form of the Lagrangian in Eq.\ (\ref{intlag})}

\label{app2}

After performing the field transformations in Eqs.\ (\ref{shift}) and
(\ref{psz}), the effective Lagrangian (\ref{intlag}) takes the form:
\begin{align}
\mathcal{L}_{\tilde{G}}^{int}  &  =\frac{c_{\tilde{G}\Phi}}{2\sqrt{2}}%
\tilde{G}(\sqrt{2}Z_{K_{S}}Z_{K}a_{0}^{0}K_{S}^{0}\overline{K}^{0}+\sqrt
{2}Z_{K}Z_{K_{S}}a_{0}^{0}K^{0}\overline{K}_{S}^{0}-2Z_{K_{S}}Z_{K}a_{0}%
^{+}K_{S}^{0}K^{-}\nonumber\\
&  -2Z_{K_{S}}Z_{K}a_{0}^{+}K_{S}^{-}K^{0}-2Z_{K_{S}}Z_{K}a_{0}^{-}%
\overline{K}_{S}^{0}K^{+}-\sqrt{2}Z_{K_{S}}Z_{K}a_{0}^{0}K_{S}^{-}K^{+}%
-\sqrt{2}Z_{K}^{2}Z_{\eta_{N}}K^{0}\overline{K}^{0}\eta_{N}\nonumber\\
&  +\sqrt{2}Z_{K_{S}}^{2}Z_{\eta_{N}}K_{S}^{0}\overline{K}_{S}^{0}\eta
_{N}-\sqrt{2}Z_{K}^{2}Z_{\eta_{N}}K^{-}K^{+}\eta_{N}+Z_{\eta_{S}}{a_{0}^{0}%
}^{2}\eta_{S}+2Z_{\eta_{S}}a_{0}^{-}a_{0}^{+}\eta_{S}\nonumber\\
&  +Z_{\eta_{N}}^{2}Z_{\eta_{S}}\eta_{N}^{2}\eta_{S}-\sqrt{2}Z_{K}^{2}Z_{\pi
}K^{0}\overline{K}^{0}\pi^{0}+\sqrt{2}Z_{K_{S}}^{2}Z_{\pi}K_{S}^{0}%
\overline{K}_{S}^{0}\pi^{0}+\sqrt{2}Z_{K}^{2}Z_{\pi}K^{-}K^{+}\pi
^{0}\nonumber\\
&  -Z_{\eta_{S}}Z_{\pi}^{2}\eta_{S}\,{\pi^{0}}^{2}+2Z_{K}^{2}Z_{\pi}%
\overline{K}^{0}K^{+}\pi^{-}+2Z_{K}^{2}Z_{\pi}K^{0}K^{-}\pi^{+}-2Z_{K_{S}}%
^{2}Z_{\pi}K_{S}^{0}K_{S}^{-}\pi^{+}\nonumber\\
&  -2Z_{\eta_{S}}Z_{\pi}^{2}\eta_{S}\pi^{-}\pi^{+}-2Z_{K_{S}}Z_{K}a_{0}%
^{-}K_{S}^{+}\overline{K}^{0}+\sqrt{2}Z_{K_{S}}^{2}Z_{\eta_{N}}K_{S}^{+}%
K_{S}^{-}\eta_{N}-\sqrt{2}Z_{K_{S}}^{2}Z_{\pi}K_{S}^{+}K_{S}^{-}\pi
^{0}\nonumber\\
&  -2Z_{K_{S}}^{2}Z_{\pi}K_{S}^{+}\overline{K}_{S}^{0}\pi^{-}-\sqrt{2}%
Z_{K_{S}}Z_{K}a_{0}^{0}K_{S}^{+}K^{-}+\sqrt{2}Z_{K}Z_{K_{S}}K^{-}K_{S}^{+}%
\phi_{N}+\sqrt{2}Z_{K}Z_{K_{S}}K^{-}K_{S}^{+}\sigma_{N}\nonumber\\
&  +\sqrt{2}Z_{K_{S}}Z_{K}K_{S}^{0}\overline{K}^{0}\phi_{N}+\sqrt{2}Z_{K_{S}%
}Z_{K}K_{S}^{0}\overline{K}^{0}\sigma_{N}+\sqrt{2}Z_{K}Z_{K_{S}}K^{0}%
\overline{K}_{S}^{0}\phi_{N}+\sqrt{2}Z_{K}Z_{K_{S}}K^{0}\overline{K}_{S}%
^{0}\sigma_{N}\nonumber\\
&  +\sqrt{2}Z_{K_{S}}Z_{K}K_{S}^{-}K^{+}\phi_{N}+\sqrt{2}Z_{K_{S}}Z_{K}%
K_{S}^{-}K^{+}\sigma_{N}-Z_{\eta_{S}}\eta_{S}\phi_{N}^{2}-Z_{\eta_{S}}\eta
_{S}\sigma_{N}^{2}-2Z_{\eta_{S}}\eta_{S}\phi_{N}\sigma_{N}\nonumber\\
&  +2Z_{\pi}a_{0}^{0}\pi^{0}\phi_{S}+2Z_{\pi}a_{0}^{0}\pi^{0}\sigma
_{S}+2Z_{\pi}a_{0}^{+}\pi^{-}\phi_{S}+2Z_{\pi}a_{0}^{+}\pi^{-}\sigma
_{S}+2Z_{\pi}a_{0}^{-}\pi^{+}\phi_{S}+2Z_{\pi}a_{0}^{-}\pi^{+}\sigma
_{S}\nonumber\\
&  -2Z_{\eta_{N}}\eta_{N}\phi_{N}\phi_{S}-2Z_{\eta_{N}}\eta_{N}\phi_{N}%
\sigma_{S}-2Z_{\eta_{N}}\eta_{N}\sigma_{N}\phi_{S}-2Z_{\eta_{N}}\eta_{N}%
\sigma_{N}\sigma_{S})\text{ .}%
\end{align}

The latter expression is used to determine the coupling of the field
$\tilde{G}$ to scalar and pseudoscalar mesons.

\subsection{Tree-body decay}

\label{app3}

For completeness we report the explicit expression for the three-body decay
width for the process $\tilde{G}\rightarrow P_{1}P_{2}P_{3}$ \cite{PDG}:%
\[
\Gamma_{\tilde{G}\rightarrow P_{1}P_{2}P_{3}}=\frac{s_{\tilde{G}\rightarrow
P_{1}P_{2}P_{3}}}{32(2\pi)^{3}M_{\tilde{G}}^{3}}\int_{(m_{1}+m_{2})^{2}%
}^{(M_{\tilde{G}}-m_{3})^{2}}dm_{12}^{2}\int_{(m_{23})_{\min}}^{(m_{23}%
)_{\max}}|-i\mathcal{M}_{\tilde{G}\rightarrow P_{1}P_{2}P_{3}}|^{2}dm_{23}^{2}%
\]
where
\begin{align}
(m_{23})_{\min}  &  =(E_{2}^{\ast}+E_{3}^{\ast})^{2}-\left(  \sqrt{E_{2}%
^{\ast2}-m_{2}^{2}}+\sqrt{E_{3}^{\ast2}-m_{3}^{2}}\right)  ^{2}\text{ ,}\\
(m_{23})_{\max}  &  =(E_{2}^{\ast}+E_{3}^{\ast})^{2}-\left(  \sqrt{E_{2}%
^{\ast2}-m_{2}^{2}}-\sqrt{E_{3}^{\ast2}-m_{3}^{2}}\right)  ^{2}\text{ ,}%
\end{align}
and%
\begin{equation}
E_{2}^{\ast}=\frac{m_{12}^{2}-m_{1}^{2}+m_{2}^{2}}{2m_{12}}\text{ , }%
E_{3}^{\ast}=\frac{M_{\tilde{G}}^{2}-m_{12}^{2}-m_{3}^{2}}{2m_{12}}\text{ .}%
\end{equation}
The quantities $m_{1},$ $m_{2},$ $m_{3}$ refer to the masses of the three
pseudoscalar states $P_{1},$ $P_{2}$, and $P_{3},$ $\mathcal{M}_{\tilde
{G}\rightarrow P_{1}P_{2}P_{3}}$ is the corresponding tree-level decay
amplitude, and $s_{\tilde{G}\rightarrow P_{1}P_{2}P_{3}}$ is a symmetrization
factor (it equals $1$ if all $P_{1},$ $P_{2}$, and $P_{3}$ are different, it
equals $2$ for two identical particles in the final state, and it equals $6$
for three identical particles in the final state). For instance, in the case
$\tilde{G}\rightarrow{K^{-}K^{+}\pi^{0}}$ one has: $|-i\mathcal{M}_{\tilde
{G}\rightarrow{K^{-}K^{+}\pi^{0}}}|^{2}=\frac{1}{4}c_{\tilde{G}\Phi}^{2}%
Z_{K}^{4}Z_{\pi}^{2}$, $m_{1}=m_{2}=m_{K}=0.494$ GeV, $m_{3}=m_{\pi^{0}%
}=0.135$ GeV, and $M_{\tilde{G}}=2.6$ GeV. Then:
\begin{equation}
\Gamma_{\tilde{G}\rightarrow{K^{-}K^{+}\pi^{0}}}=0.00041\,c_{\tilde{G}\Phi
}^{2}\text{ [GeV]}\;. \label{4}%
\end{equation}
The full decay width into the channel $KK\pi$ results from the sum%
\begin{equation}
\Gamma_{\tilde{G}\rightarrow{KK\pi}}=\Gamma_{\tilde{G}\rightarrow{K^{-}%
K^{+}\pi^{0}}}+\Gamma_{\tilde{G}\rightarrow{K^{0}\bar{K}^{0}\pi^{0}}}%
+\Gamma_{\tilde{G}\rightarrow{\bar{K}^{0}K^{+}\pi^{-}}}+\Gamma_{\tilde
{G}\rightarrow{K^{0}K^{-}\pi^{+}}}=6\Gamma_{\tilde{G}\rightarrow{K^{-}K^{+}%
\pi^{0}}}\text{ .}%
\end{equation}
The other decay channels can be calculated in a similar way.

\end{document}